\documentclass[journal]{IEEEtran}
\usepackage{soul,color}
\usepackage[T1]{fontenc}
\usepackage{graphicx} 
\graphicspath{ {./images/} }
\usepackage{algorithm}
\usepackage{algpseudocode}
\usepackage{amsmath}
\usepackage{amssymb}
\usepackage{mathpazo}

\ifCLASSINFOpdf
\else
\fi
\usepackage[caption=false,font=footnotesize]{subfig}
\begin{document}
%
\title{E-cheating Prevention Measures: Detection of Cheating at Online Examinations Using Deep Learning Approach - A Case Study}
%
%
%

\author{Leslie~Ching~Ow~Tiong and
        HeeJeong~Jasmine~Lee
\thanks{L.C.O. Tiong is with Computational Science Research Center in Korea Institute of Science and Technology (KIST), Seoul 02792, Republic of Korea (email: tiongleslie@kist.re.kr).}
\thanks{H.J. Lee is with Pierson College in PyeongTaek University, Pyeongtaek-si, Gyeonggi-Do 17869, Republic of Korea (email:hjlee@ptu.ac.kr).}
}

%
%

\markboth{Journal of \LaTeX\ Class Files,~Vol.~XX, No.~XX, Jan~2021}%
{Shell \MakeLowercase{\textit{et al.}}: Bare Demo of IEEEtran.cls for IEEE Journals}
%



\maketitle

\begin{abstract}
This study addresses the current issues in online assessments, which are particularly relevant during the Covid-19 pandemic. Our focus is on academic dishonesty associated with online assessments. We investigated the prevalence of potential e-cheating using a case study and propose preventive measures that could be implemented. We have utilised an e-cheating intelligence agent as a mechanism for detecting the practices of online cheating, which is composed of two major modules: the internet protocol (IP) detector and the behaviour detector. The intelligence agent monitors the behaviour of the students and has the ability to prevent and detect any malicious practices. It can be used to assign randomised multiple-choice questions in a course examination and be integrated with online learning programs to monitor the behaviour of the students. The proposed method was tested on various data sets confirming its effectiveness. The results revealed accuracies of 68\% for the deep neural network (DNN); 92\% for the long-short term memory (LSTM); 95\% for the DenseLSTM; and, 86\% for the recurrent neural network (RNN).
\end{abstract}


%
\IEEEpeerreviewmaketitle

\section{Introduction}
\label{sec::sec1}
%
%
%
%
\IEEEPARstart{O}{nline} courses have become a feasible option in education. This platform is increasingly recognised in colleges and higher education institutions such as universities, and even implemented in elementary schools – for example, during the Covid-19 pandemic. However, the detached nature of online education raises concerns about the potential risks of academic dishonesty, particularly when students sit for exams at remote locations, in the absence of the disciplinary procedures that are typically employed at examination centres \cite{Sarrayrih2013, Jalali2017, Raud2019}. There is exponential growth in online education, in terms of both student enrolment and the corporate market it entails. However, existing literature indicate a prevalence of online cheating, which involves academic dishonesty by both the faculty and the students \cite{Rowe2004, Moten2013, Chuang2017}. Although online education provides valuable learning opportunities for people who do not have access to traditional quality education due to time or physical constraints, its credibility may be compromised if issues of academic dishonesty are not resolved.

Although online courses have increasingly gained momentum during the Covid-19 pandemic, it should not be established as an accepted mode of education without due diligence in the event of a possible resurgence of the pandemic. For instance, the detached nature of online courses has raised significant concerns as prevalence of e-cheating has been reported \cite{Chuang2017}. This is because, while traditional sit-down examinations are invigilated, the same cannot be said for the remotely conducted online examinations. Consequently, the credibility of online courses could become questionable.

The present study aims to address the current limitations of cheating at online examinations by proposing artificial intelligence (AI) techniques via the internet protocol (IP) network detector and deep learning-based behaviour detection agent. The research was conducted as a case study, the outcomes of which offer avenues of further improving the intelligent tutoring system. The main contributions of this study are as follows:
\begin{itemize}
\item We analysed the limitations of the current online education system, with particular focus on cheating at online examinations.
\item We proposed an e-cheating intelligence agent that is based on the relationship model for detecting online cheating using AI techniques. Specifically, we implement an IP detector and a behaviour detector that utilises the long short-term memory (LSTM) network with a densely connected concept, namely DenseLSTM. As AI techniques have evolved rapidly, and has been widely applied in recent years, we utilised state-of-the-art AI techniques for the online exams, which may provide useful insights that contribute to the research area of an intelligent tutoring system.
\item We created a new dataset for this study, which is presented in \cite{ourdb2020}. The records were collected across online exams that were conducted in environments that were highly unregulated (e.g., in the absence of any implemented applications for detecting or preventing e-cheating) during mock, mid-term, and final-term exam periods. The database included training and testing schemes for performance analysis and evaluation.
\end{itemize}

The paper is organised as follows: Section \ref{sec::sec2} analyses the limitations of the existing online education system for preventing online cheating. The proposed framework is presented in Section \ref{sec::sec3} and the detailed database information is presented in Section \ref{sec::sec4}. Section \ref{sec::sec5} discusses the experimental results, and the conclusions are summarised in the last section.


\section{Background and Related Works}
\label{sec::sec2}
Students engaging in cheating during examinations is a prevalent phenomenon worldwide, regardless of the country's stage of development. To this end, early work done by Roger \cite{Roger2006}, suggested installing the proctoring security program into computers, which enables continuous monitoring of each student’s computer screen on an instructor control view. Similarly, Cluskey \textit{et al.} \cite{Cluskey2011}, proposed an eight-step model for reducing potential cheating by students, which incorporated several criteria that were contributory to the exam, such as the duration of the exam, the time allocated for answering each question, etc. However, this model has limitations given that it required the students to use a special browser in order to access the examination applications, and the consequent necessity for the teacher to change the questions every semester by using the randomised approach \cite{Cluskey2011}.

\subsection{Network Security Methods}
The rapid growth of wireless communication technology enables users to remotely access digital resources at any time. Many researchers have focused their work on strengthening the security of online exam protocols to counter the security breaches that could occur in the education sector. Bella \textit{et al.} \cite{Bella2015} proposed a new protocol for online examinations, which bypassed the necessity of involving a trusted third party to maintain discipline. The proposed protocol merged oblivious transfer and visual cryptography that allowed both the students and invigilators to generate aliases. These would only be revealed during the exam, thereby maintaining the anonymity of the students.

A literature survey by Ullah \textit{et al.} \cite{Ullah2016} discusses the security threats that have been encountered in the past, which are associated with online examinations. They indicated collusion to an be increasingly challenging threat, which typically involves the collaboration of a third party who assists the student by impersonating him or her online. A further study conducted by the same authors \cite{Ullah2018} revealed the potential mechanisms of security attacks in online cheating. They monitored 31 online participants at an examination, where they assessed the students' behaviour by employing dynamic profile questions in an online course. The results pointed out that students who cheated by impersonation shared most of the information using a mobile device, and consequently, their response time was significantly different to those who did not cheat.

A recent literature review \cite{Fernandes2018} highlights the significance of advanced technology in enabling the use of techniques such as anomaly detection for addressing the growing concerns of e-cheating. The authors reviewed the current work under five dimensions: network data type, network traffic, intrusion detection, detection methods and open issues. They also emphasised the accurate identification of an individual to be crucial in developing any intrusion detection system that is aimed at restricting their access within the network traffic.

\subsection{Plagiarism Detection Methods and Tools}
Plagiarism detection tools are popular in course evaluations for identifying the unpermitted use of written content by students. For instance, the code plagiarism tool can be used to calculate the similarity between a pair of programs using a token sequence \cite{Mariani2012} and dependency graph features \cite{Chen2014}. Nonetheless, these code plagiarism detection techniques can be circumvented by modifying the code syntax. In order to counter such deceit, Herrera \textit{et al.} \cite{Herrera2019} implemented a new language-agnostic methodology, which prevents plagiarism in programming courses without the necessity for code comparison or professor intervention.

A study carried out by Pawelczak \cite{Pawelczak2018} examined the achievements and opinions of students over five years regarding the automated evaluation system used in plagiarism detection in programming courses. The data comprised of 228 records, where the analysis was based on tokenizing and averaging several features of the source code. The study revealed a limitation in the method: given that plagiarism detection is dependent on the thresholding value that is used to calculate the level of similarity in the content, thresholding issues in this approach could give rise to false measures of prevention.

\subsection{Biometrics Methods}
It is vital to ensure the presence of an examinee throughout the entire examination. The invigilation of online examinations is difficult, and the absence of a physical invigilator gives rise to a higher possibility of suspicious conduct as well as cheating attempts taking place. Several strategies have been proposed to counter such fraudulent activities during examinations including the monitoring of yaw angle variations, audio presence and active window capture. For instance, Prathish \textit{et al.} \cite{Prathish2016} and Narayanan \textit{et al.} \cite{Narayanan2014} proposed to implement the features of point extraction and yaw angle detection that could assist the instructors in monitoring the students during online examinations. Similarly, Wlodarczyk \textit{et al.} \cite{Wlodarczyk2016} presented the head pose detection method, which uses the precise localisation of face landmark points that help in identifying the user’s direction of gaze as well as facial recognition.

 Hu \textit{et al.} \cite{Hu2018} proposed a novel method of monitoring the students' behaviour during online assessments, which involved identifying the relevant relationship between the image of a student's face and his or her corresponding pose. Mahadi \textit{et al.} \cite{Mahadi2018} suggested that the combination of facial recognition and keystroke dynamics could be the best classifiers for behavioural-based biometric authentication. In similar a study, Ghizlane \textit{et al.} \cite{Ghizlane2019} primarily focused on the identity and access management of students and staff, where they could use a customised model of a smart card-based digital identity control system for accessing academic services, especially online evaluations. This service would help in ensuring that the security of the institute could manage the authorisation of individuals in accessing various types of networks.

Ghizlane \textit{et al.} \cite{Ghizlane2019b} also examined the combined use of smart cards and face recognition to authorise and monitor applicants while online exams are conducted in detecting any suspicious behaviour or cheating attempts. They recommended a system that stored a log of photographs taken of each applicant that sits the exam, which could be later checked by administrators. Another study conducted by Garg \textit{et al.} \cite{Garg2020} proposed using the convolutional neural network and the Haar cascade classifier to detect the faces of the exam candidates and to tag these with an associated name that is given at the time of registration, which allows the system to keep track of the applicants’ movements within the timeframe of the exam.

In current online examination settings, biometric authentication is considered to be one of the most popular techniques in verifying the identification of the candidates \cite{Fayyoumi2014, Karim2015}. In contrast to face-to-face examinations, online assessments do not involve proctors or invigilators, and can be held in different and uncontrolled remote environments. Consequently, establishing authentication goals in online exams are vital in order to verify the identity of the students, as it plays a key role in online security \cite{Bawarith2017, Hadian2019}. In a study which focused on enhancing the security of online examinations, Mathapati \textit{et al.} \cite{Mathapati2017} proposed utilising personal-images as graphical passwords. They suggested using digital pictures that were captured from live video as personalised physical tokens.

Ramu \textit{et al.} \cite{Ramu2013} and Mungai \textit{et al.} \cite{Mungai2017} reviewed the importance of keystrokes dynamics in preserving security in online examinations. The proposed architecture used a three-stage authentications process, in which the stages were described as statistical, machine learning and logic comparison. Initially, when an applicant signs into the system, his/her typing style is automatically recorded, for which a template is generated. These templates are subsequently used as a guide to continuously monitor the authenticity of the users, based on several parameters: their dwell time (the time difference between keypress and release); the flight time (the time passed between the key release and keypress of two consecutive keystrokes); and, the typing speed, for better precision and robustness. Ananya and Singh \cite{Ananya2018} also introduced the keystroke dynamics approach, in which the system does not require any pre-registration and has the capability of keeping track of each student’s typing pattern during the exercise session itself.

\subsection{Summary}
Due to the Covid-19 pandemic, many academic institutes, schools and universities have switched to online teaching, and consequently, online examinations have become a common trend, especially due to its flexibility and usability within different environments \cite{Almaiah2020, Rajab2020, Adedoyin2020}. Several studies offer diverse approaches to mitigate cheating at online examinations that largely focus on behaviour analysis, technology innovation, etc. \cite{Shukor2015, Yan2019, Gonzalez2020}. However, the detection of suspicious behaviour of candidates at online examinations still remains one of the major challenges in fully utilising online education platforms. In this context, we offer a solution for detecting abnormal behaviour of students at online examinations, and thereby for preventing e-cheating, by investigating the use of an AI intelligence agent as a real-time live proctor. The AI intelligence agent was designed utilising network protocol detection and deep learning approaches.

\section{Methodology}
\label{sec::sec3}
\begin{figure*}[!t]
\centering
\includegraphics[width=0.98\textwidth]{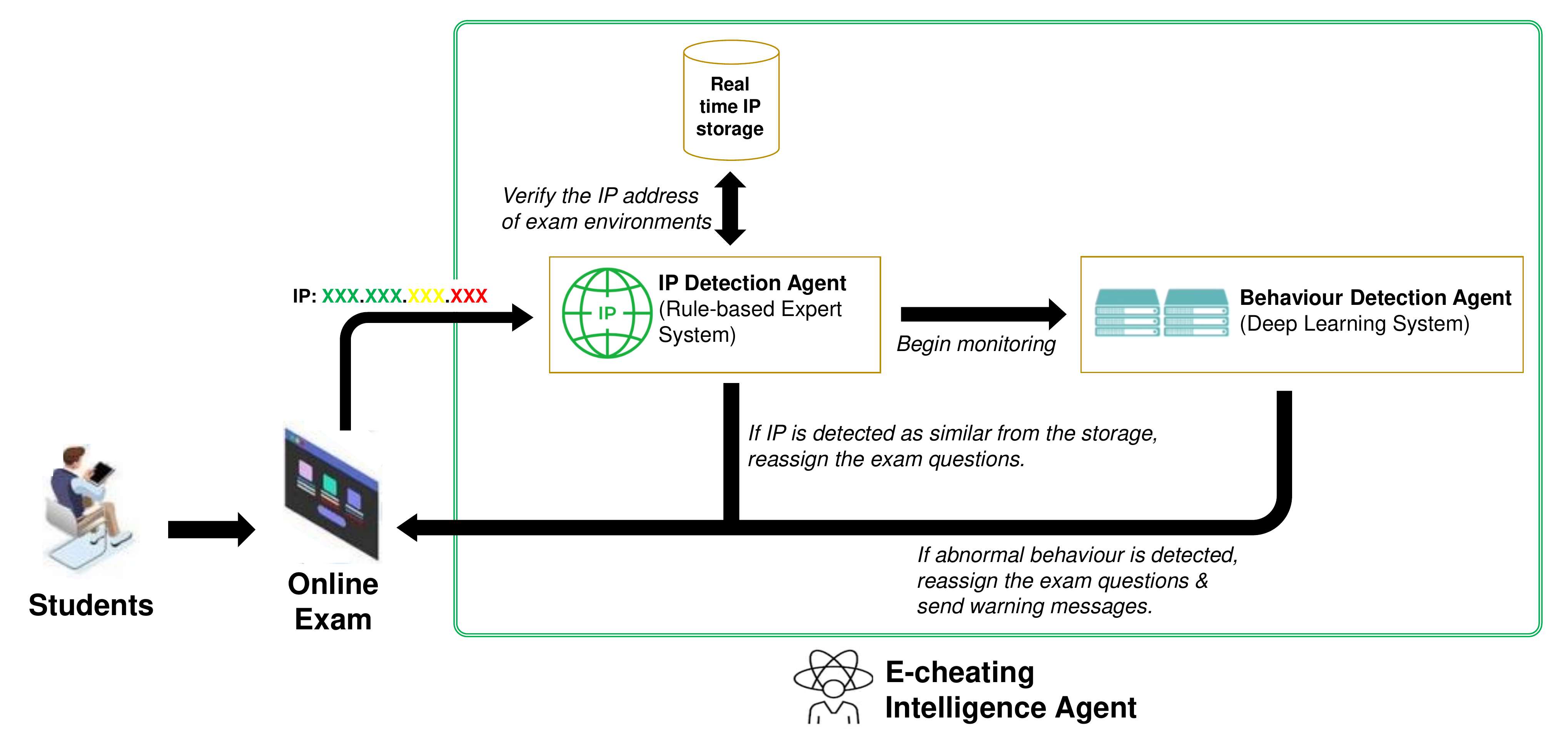}
\caption{The proposed framework for the e-cheating intelligent agent system. This framework describes the entire process for applying the e-cheating intelligent agent to identify abnormal behaviour and to prevent e-cheating at online examinations.}
\label{fig::fig1}
\end{figure*}

We designed an online examination as a case study, which consisted of multiple-choice questions, in which an e-cheating intelligent agent was used to detect any potential cheating. The e-cheating intelligence agent consists of two main agents: the network IP detection agent (described in Section \ref{subsec::subsec31}) and the behaviour detection agent (described in Section \ref{subsec::subsec32}). Fig. \ref{fig::fig1} illustrates the architecture of the proposed system.

\subsection{Network IP Detection Agent}
\label{subsec::subsec31}
Emerging security analysis has raised awareness of the challenges of online learning and has captured the rapidly increasing attention of researchers of developing new e-learning assessment methods. The present issues arises from the inadequate understanding of the security dataset that is stored through network protocols and the analyses of data that use semantic association and inference methods \cite{Yao2016}.

The proposed model is a two-stage process (see Fig. \ref{fig::fig1}). In the first stage, we propose applying an IP detection agent to filter any deceitful activity. For example, the system can monitor the exam candidates' IP addresses. Most routers allocate dynamic IP addresses, which are numerical labels that are specifically assigned to each device that is connected to a computer network. This would enable the system to issue an alert if a student changed their computer device or their initial location. In the proposed method, there would be several sets of exam questions (such as Set A, B, C, etc.). At the start of an examination, after verification, a student is randomly assigned a set of questions for assessment (e.g., Set A). If any abnormal behaviour is detected, the system changes the questions randomly to another Set (e.g., Set B). Similarly, if an incoming student enters his/her credential to access the online exam platform using an IP address that was previously recorded in the IP list as suspicious, the system will generate a different set of questions to the student. Algorithm 1 summarises the entire process of the IP detection agent.

\begin{algorithm}
\caption{IP detection agent}
\label{alg::alg1}
\textbf{Input:} IP address from a new examinee $E$ \newline
\textbf{Output:} Decision $D$ \newline \newline
\textbf{//Initialisation} \newline
Let $\text{IP}_{DB}$ be the list of real-time IP with $N$ size \newline \newline
\textbf{if} $E$ is not in $\text{IP}_{DB}$ \textbf{then}
\begin{algorithmic}
\State Add $E$ into $\text{IP}_{DB}$
\State \textbf{return} $D$ with random sets to $E$
\end{algorithmic}
\textbf{else}
\begin{algorithmic}
\State \textbf{for} $i \rightarrow$ 1 to $N$ \textbf{do}
\State $\;\;\;$ \textbf{if} $E$ is found in $\text{IP}_{DB}[i]$ \textbf{then}
\State $\;\;\;\;\;\;\;$ \textbf{return} $D$ with specific sets to $E$
\State $\;\;\;$ \textbf{end}
\State \textbf{end}
\end{algorithmic}
\textbf{end}
\end{algorithm}

\subsection{Behaviour Detection Agent}
\label{subsec::subsec32}
We devised a behaviour detection agent via a deep learning approach to monitor and analyse the behaviour of all the students. As illustrated in Fig. \ref{fig::fig1}, the agent would alert the instructors and immediately reassign the remaining questions with a new set of questions only in instances where abnormal behaviour is detected in the students during the examinations. The following sub-sections provide a more detailed explanation of the behaviour detection agent.

\subsubsection{Data pre-processing} 
\begin{figure*}[!t]
\centering
\includegraphics[width=0.98\textwidth]{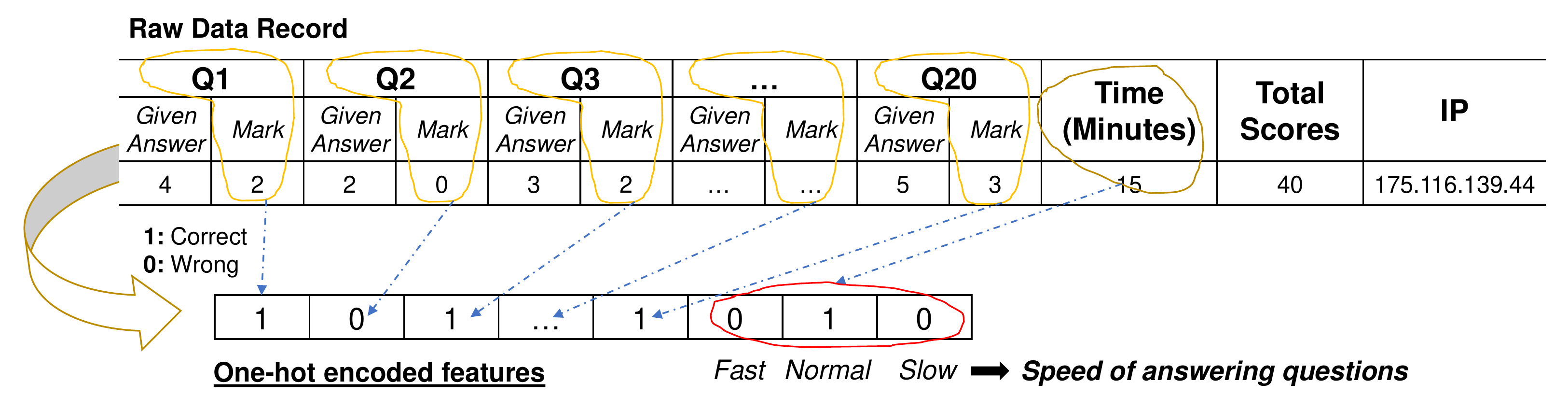}
\caption{Generation of the one-hot encoded feature. The example demonstrates the process of generating one-hot encoded features from our dataset.}
\label{fig::fig2}
\end{figure*}

Before training the behaviour detection agent, we first transformed each raw data record into a one-hot encoded feature, which defines the behaviour of the student during the examination. For instance, each raw data record contains the results of the 20 multiple-choice questions, the total time (in minutes) taken for answering the examination, and the final score. In this study, we defined the one-hot encoded input feature as: $R \in 1 \times N$, where $N=23$. The first 20 elements represent the given answers of the 20 questions as: $[1, 1, 1, 1, 1, 0, 1, 0, \cdots, 0]$ where the values 1 and 0 reflect whether the answer is correct or incorrect, respectively. The last three elements define the speed of answering the questions, whether fast, normal or slow. The last three elements are defined as the speed of answering questions as fast, normal or slow. Fig. \ref{fig::fig2} summarises data pre-processing, where the raw data is processed into one-hot encoded features.

Next, we explored how we could label the data record to identify the students' behaviours. Initially, we labelled the records as one of two major categories of behaviours: `normal' or `abnormal'. In defining ‘abnormal’ behaviour, we assessed the speed at which the students have answered in instances where the questions have been answered 90\% correctly, according to one-hot encoded features. If the speed was represented as too fast or too slow, they were labelled as `abnormal'; in comparison, the rest of the samples were considered as `normal'.

When defining the speed of answering questions two factors were considered: the number of questions answered and the level of difficulty of each question. For example, we observed that if the level of difficulty of a question was defined as `easy', most of the students could answer it within 10 to 20 seconds; if it was defined as `moderate' or `high', they required 30 to 40 seconds or 1-2 minutes, respectively. However, the specifics of such labelling criteria are dependent on the subjects or the courses that are evaluated.

\subsubsection{DenseLSTM} We propose applying a deep learning network, namely DenseLSTM as the behaviour detection agent. The LSTM network was introduced by  Hochreiter and Schmidhuber \cite{Hochreiter1997}, and allows modelling the problem with sequential dependencies data, such as time-series data, natural language processing, behaviour analysis, etc. In our work, we propose to use a densely connected approach with the LSTM network to extract better feature representation for abnormal behaviour prediction.

\begin{figure}[!t]
\centering
\includegraphics[width=0.47\textwidth]{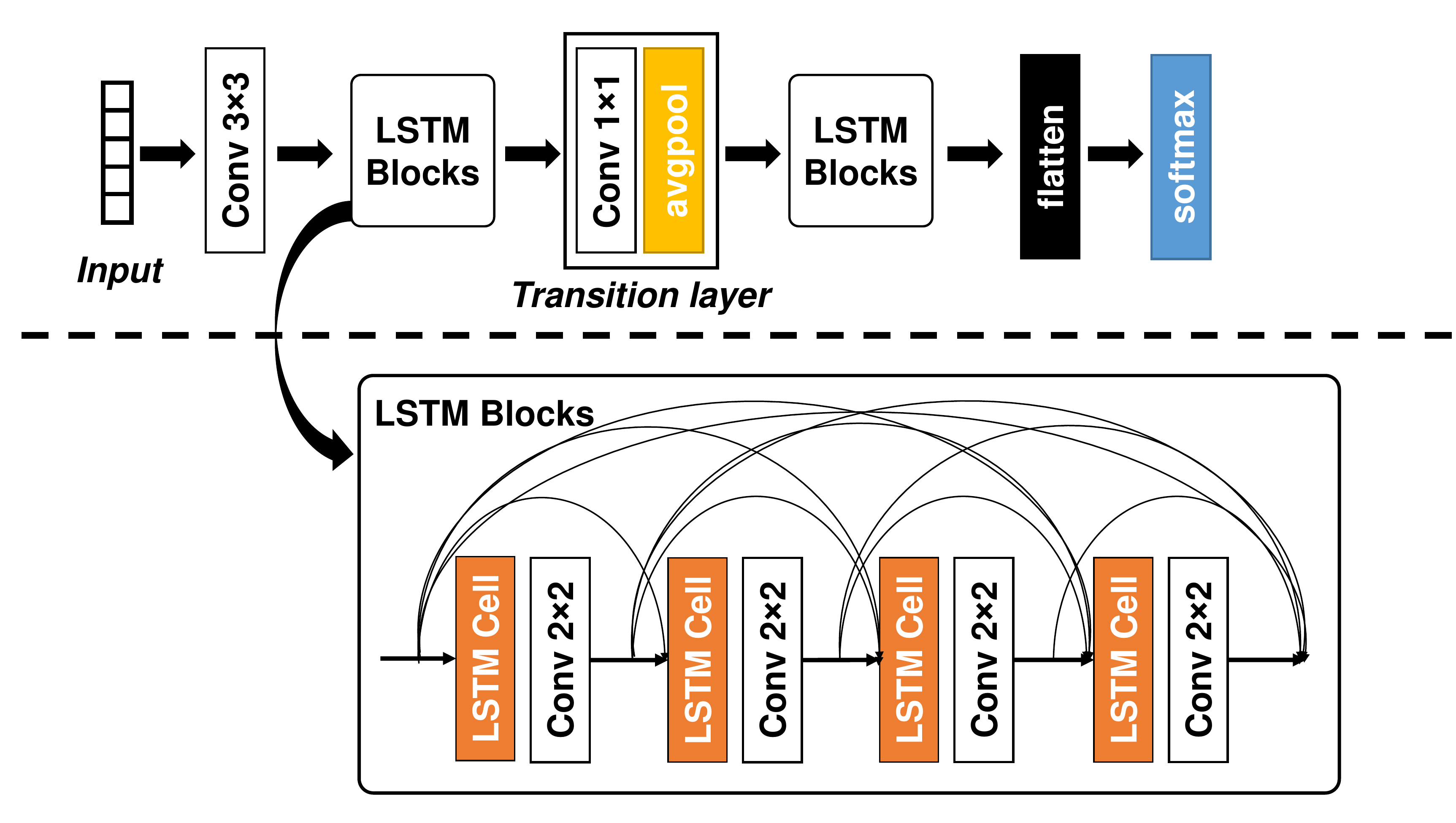}
\caption{Architecture of the DenseLSTM.}
\label{fig::fig3}
\end{figure}

Fig. \ref{fig::fig3} illustrates the architecture of the DenseLSTM, which consists of a convolutional (\textit{conv}) layer, two LSTM blocks and a transition layer. The concept of the densely connected network was originally proposed by Huang \textit{et al.} \cite{Huang2019} for image classification. The network introduces direct connections from any layer to all the subsequent layers, which improves the information flow between them by creating a different connectivity pattern. One explanation for this occurrence is the enhanced access each layer has to all the preceding feature maps in its block due to the dense connectivity, and the ``collective knowledge'' it thereby provides the network. 

\begin{table}[!t]
\renewcommand{\arraystretch}{1.2}
\caption{The configurations of each layer for the proposed network. $f$ refers to the size of feature maps and $k$ refers to the size of filter.}
\label{tab::tab1}
\centering
\begin{tabular}{lc}
\hline
\hline
\textbf{Network Layer} & \textbf{Configuration} \\ \hline
$conv_{1}$ & \textit{f}: 64$@$1$\times$23; \textit{k}: 1$\times$2; stride 2 \\
$LSTM\_Block_{1}$ & $\begin{bmatrix}
LSTM\_Cell\\ 
2\times 2 \:\: conv
\end{bmatrix} \times 4$ \\
$transition$ & 1$\times$1 $conv$; $avgpool$; stride 2 \\
$LSTM\_Block_{2}$ & $\begin{bmatrix}
LSTM\_Cell\\ 
2\times 2 \:\: conv
\end{bmatrix} \times 8$ \\
$LSTM\_Cell$ & 512$@$1$\times$6 \\
$flatten$ & 1$\times$6$\times$512 \\
\textbf{Y} & 1$\times C$ \\ 
\hline
\hline
\end{tabular}
\end{table}

In the LSTM block, the LSTM cell layer decides what information will be discarded from the cell state. As the layer could potentially inadvertently omit useful information, we have implemented the concept of a densely connected network, which keeps the information together instead of deciding ``what to forget and which new information should be added.'' The features can be accessed from anywhere within the network, and unlike in traditional network architectures, there is no requirement to replicate them from layer to layer. The network used in this study is expected to collect rich information while maintaining a low complexity of features, which can result in achieving a better classification performance.

\begin{table*}[!t]
\caption{Ten sample records of mock exam period from our dataset.}
\label{tab::tab2}
\centering
\begin{tabular}{ccccccccccc}
\hline
\hline
\textbf{ID} & \textbf{Q1 Ans.} & \textbf{Q1 Score} & \textbf{Q2 Ans.} & \textbf{Q2 Score} & \textbf{$\cdots$} & \textbf{Q20 Ans.} & \textbf{Q20 Score} & \textbf{Grade} & \textbf{Time} & \textbf{IP}\\ \hline
2000001 & 4 & 2 & 2 & 2 & $\cdots$ & 4 & 3 & 45 & 15 & 175.116.139.44 \\
2000002 & 4 & 2 & 2 & 2 & $\cdots$ & 4 & 3 & 36 & 20 & 211.214.126.62 \\
2000003 & 4 & 2 & 2 & 2 & $\cdots$ & 1 & 0 & 38 & 20 & 125.186.174.50 \\
2000004 & 1 & 0 & 2 & 2 & $\cdots$ & 2 & 0 & 24 & 18 & 1.236.192.19 \\ 
2000005 & 4 & 2 & 2 & 2 & $\cdots$ & 2 & 0 & 33 & 17 & 58.236.177.182 \\ 
2000006 & 3 & 0 & 2 & 2 & $\cdots$ & 2 & 0 & 24 & 14 & 180.71.78.211 \\ 
2000007 & 4 & 2 & 2 & 2 & $\cdots$ & 4 & 3 & 46 & 25 & 211.243.246.3 \\ 
2000008 & 4 & 2 & 2 & 2 & $\cdots$ & 4 & 3 & 45 & 24 & 211.243.246.3 \\ 
2000009 & 4 & 2 & 2 & 2 & $\cdots$ & 1 & 0 & 35 & 26 & 221.147.167.237 \\ 
2000010 & 3 & 0 & 2 & 2 & $\cdots$ & 1 & 0 & 23 & 14 & 61.74.229.32 \\ 
\hline
\hline
\end{tabular}
\end{table*}

Let $B$ denote the LSTM block with $l$ in $H$ layers, composed of LSTM cell, $conv$ layer, rectified linear unit (ReLU) \cite{Nair2010} and dropout layers: 
\begin{equation}
B = H_{l}([x_{0}, x_{1}, x_{2}, \cdots, x_{l-1} ]),
\label{eq::eq1}
\end{equation}
where $x_{0}$ to  $x_{l-1}$ represent feature outputs and $[\cdot]$ is defined as a concatenation operator. We define $l$ as 4 in the first block and $l$ as 8 in the second block. A transition layer is implemented in the first block that performs $1\times 1$ $conv$ and $avgpool$ operations, where $1\times 1$ $conv$ is defined as the filter size of the $conv$ layer that is $1\times 1$. Table \ref{tab::tab1} tabulates the architecture of the proposed network.

In the training stage, we implement a softmax cross-entropy $\mathcal{L}$ of logit vector and the respective encoded label:
\begin{equation}
\mathcal{L}(\textbf{Y}) = - \sum_{i}^{E} \sum_{j}^{C} L_{ij} \textnormal{log} ( \textnormal{softmax}(\textbf{Y})_{ij}),
\label{eq::eq2}
\end{equation}

\begin{equation}
\textnormal{softmax} ( \textbf{Y} )_{ij} = \frac{\textnormal{exp}^{\textbf{Y}_{ij}}}{\sum_{j}^{C} \textnormal{exp}^{\textbf{Y}_{ij}}},
\label{eq::eq3}
\end{equation}
where $L$, $E$ and $C$ denote class labels, the number of training samples in $\textbf{Y}$ and the number of classes, respectively.

In summary, we have devised a network using the concept of a densely connected approach to extract better feature representation and to strengthen the feature activation of the network for predicting potential e-cheating.

\section{Dataset}
\label{sec::sec4}
We have designed a new dataset, namely, the 7wiseup behaviour dataset, which consists of 94 student records that were acquired from different end-of-term examinations at Pyeongtaek University in South Korea. All records were collected during the Spring semester, 2020.

\subsection{Collection Setup}
\label{subsec::subsec41}
In order to create our database, we utilised an e-class learning management system (LMS), which is adopted by the university. The exam module is crucial for the LMS system, which can be used to set up quizzes and exam questions. The module allows the lecturers to fill out forms that outline vital information such as the exam name, time allocations (opening time, closing time and length of the exam), grades, whether shuffling of questions is allowed, etc. They can add questions from the question bank, using various in-built options for quiz/test settings including the type of examination questions – whether multiple-choice, true/false and short answers, and also allocate marks for each question. The system further enables the staff to add resources including images or links, and to provide general feedback. Once the exam is completed and submitted, the system automatically assigns marks for the questions based on the answers that were pre-determined by the staff. Finally, the system generates a CSV file compiling a list of all the submissions and the candidates’ records including their names, IDs, the answers for each question, the scores for each question, the grades, the total time taken for completion (in minutes), and the IP addresses. Table \ref{tab::tab2} shows several records of our dataset.

\subsection{Training and Benchmarking Protocols}
\label{subsec::subsec42}
For training and benchmarking protocols, 94 records each were obtained from mock-exams, and the mid- and final-term exams, respectively. Notably, the records that were used for the training and benchmarking protocols did not overlap. When designing the protocol to develop or train our model, we divided the data set from mock-exams for training and cross-validations at a ratio of 80:20. In addition, due to the imbalanced nature of the dataset, we have applied a data augmentation approach that generated an additional 60 samples representing cases of abnormal behaviour. In the benchmarking scheme, the task was to determine the behaviour of the examinees based on their manner of answering questions.

\section{Experiments}
\label{sec::sec5}
We conducted several experiments to evaluate the relative performance between our network and other benchmark networks. All the configurations used for the networks are described in Section \ref{subsec::subsec51} and the experimental results are presented in Section \ref{subsec::subsec52}.

\subsection{Experimental Setup}
\label{subsec::subsec51}

\subsubsection{Configuration of DenseLSTM}
The proposed network was implemented using TensorFlow \cite{TensorFlow}. For the configuration, we applied a learning rate of $1.0\times 10^{-5}$ and the AdamOptimizer \cite{Kingma2015}, where the weight decay and momentum were set to $1.0\times 10^{-4}$ and 0.9, respectively. In the experiments, the batch size was set to 32 and the training was carried out across 250 epochs. The training was conducted using our database according to the protocols set out in Section \ref{subsec::subsec42}; it was performed by the NVidia RTX 2080 Ti GPU.

\begin{figure*}[!t]
\centering
\subfloat[Mid-term examination]{\includegraphics[width=3.5in]{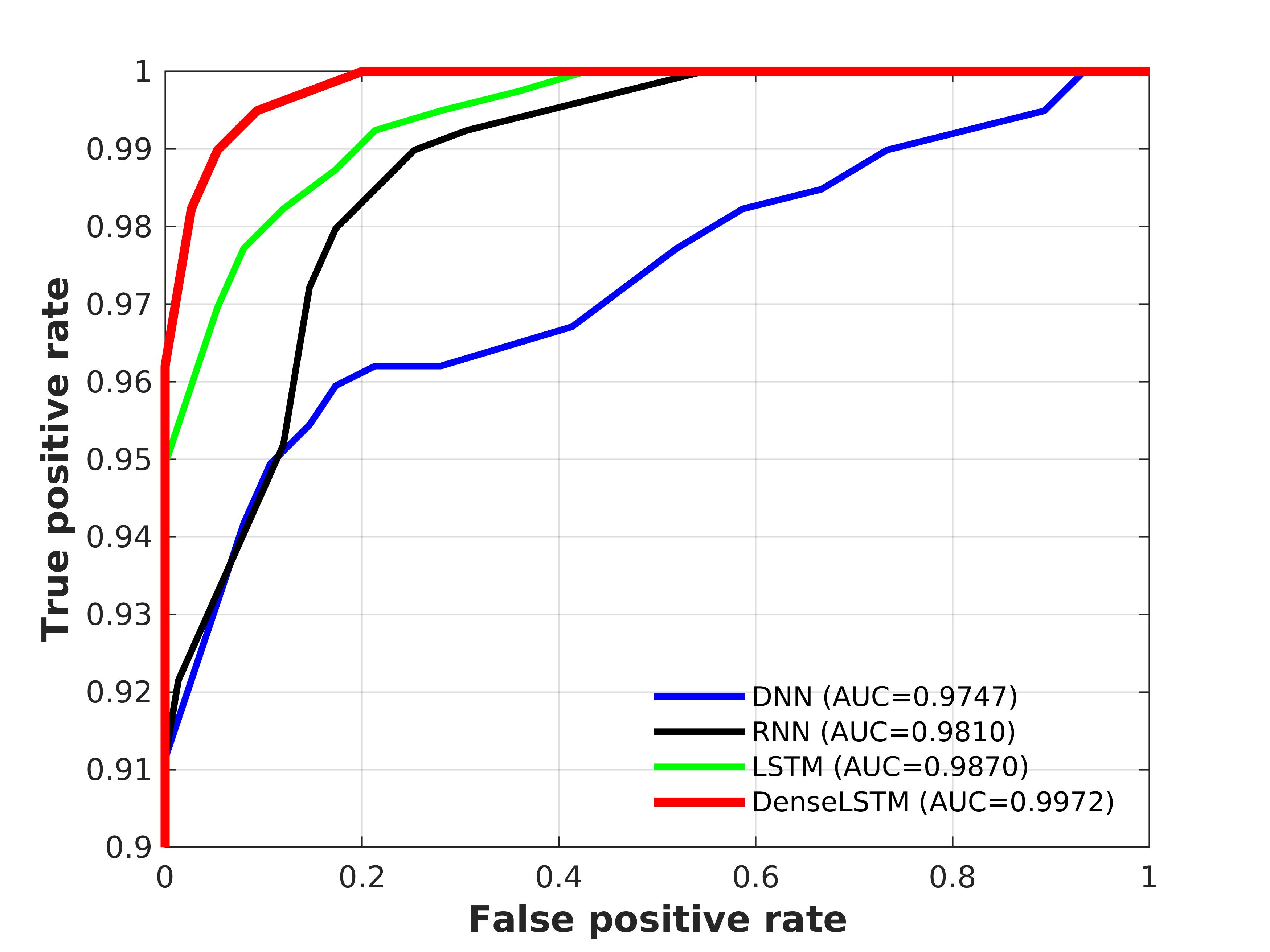}%
\label{fig::fig4a}}
\hfil
\subfloat[Final-term examination]{\includegraphics[width=3.5in]{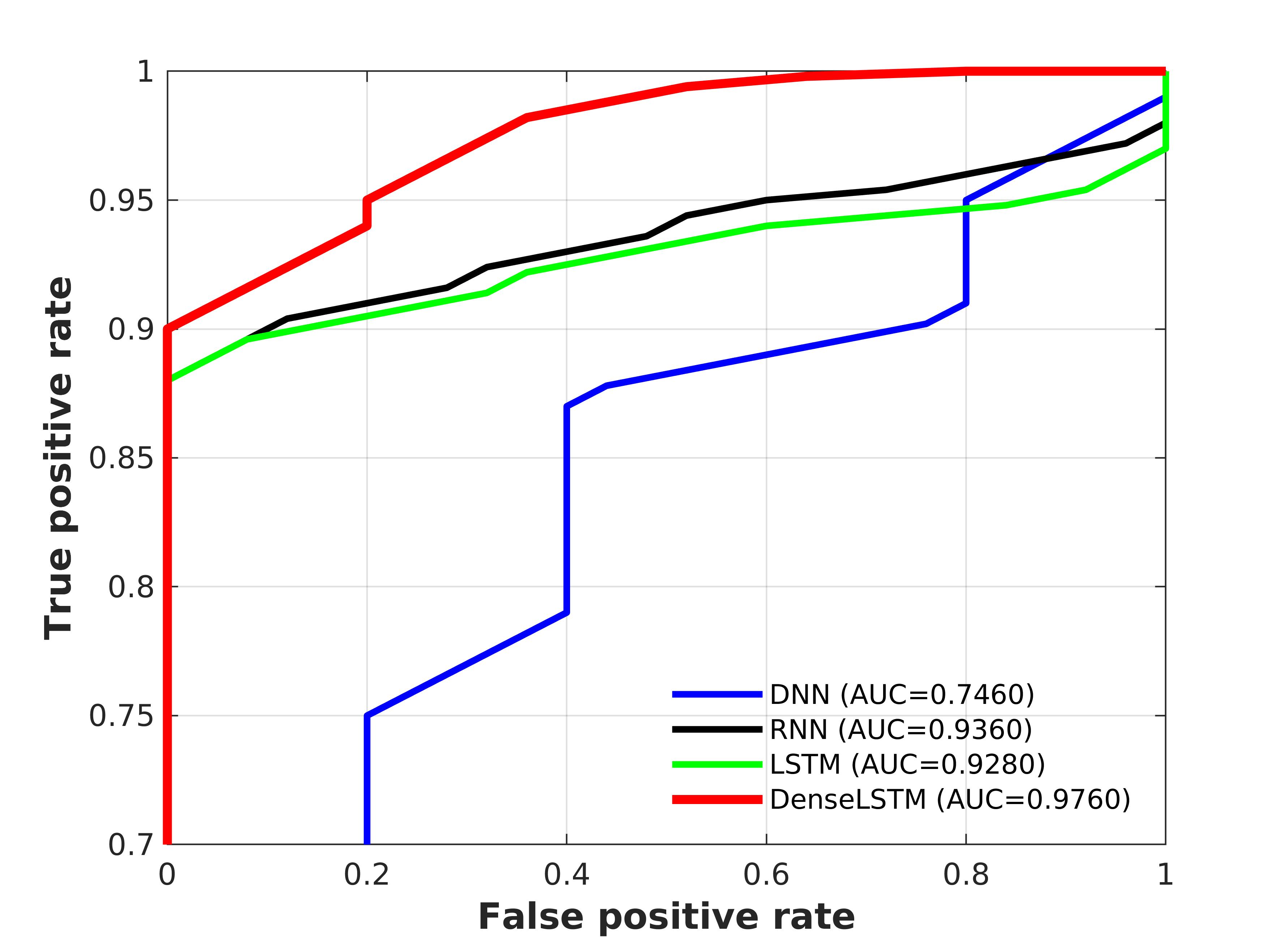}%
\label{fig::fig4b}}
\caption{Performance comparisons of the ROC curve for mid-term and final-term examinations.}
\label{fig::fig4}
\end{figure*}

\subsubsection{Configuration of Benchmark Networks}
We selected several deep networks to evaluate the performance of the behaviour study: the Deep Neural Network (DNN) \cite{Kriegeskorte2019}, the LSTM \cite{Greff2017} and the Recurrent Neural Network (RNN) \cite{Rumelhart1986}. These approaches have been successfully applied in behaviour studies previously \cite{Zerkouk2019} and \cite{Hao2019}. When conducting the experiments, we tried our best to implement and fine-tune the LSTM and RNN from scratch following the recommendations of \cite{Greff2017} and \cite{Rumelhart1986}, respectively. The training was also conducted using our database and following the protocols in Section \ref{subsec::subsec42}. All the training processes were performed using the NVidia RTX 2080 Ti GPU.

\subsection{Experimental Results}
\label{subsec::subsec52}
In order to evaluate the performance of our network in a real-life scenario and in a more objective manner, we used the mid-term and final-term examinations data and compared our experimental results with the results obtained from other benchmark approaches. As shown in Table \ref{tab::tab3}, of the different approaches tested, the highest accuracy, of 95.32\%, was achieved by our system for overall performance, followed by the second-best, the LSTM system, which reached an accuracy of 91.89\%. Our model outperformed the existing benchmark models by 3.5\% in accuracy demonstrating its superiority to the other benchmark approaches.

In identifying the source of the improvements in our system, we observed that the accuracy of our network was higher than 92\% in analysing the behaviour of students during online exams for both mid-term and final-term examinations, with values of 97.77\% and 92.86\%, respectively (as shown in Table \ref{tab::tab3}). The LSTM system, which was the second-best, demonstrated comparable accuracy scores of 94.49\% for the mid-term and 89.29\% for the final-term examinations. In contrast, the performance of the DNN approach was low with an accuracy score of 82.74\% for the mid-term examination, and only 52.68\% accuracy for the final-term examination. In addition, our investigations of the DNN system revealed error rates of 2.23\% and 7.14\% for mid-term and final-term examinations, respectively, which was associated with false prevention that occurred as a result of several students being extremely slow to select their answers.

With the intention of refining our system further, we focused on characterising its sensitivity and specificity by applying the parameters, Receiver Operating Characteristics (ROC) and Area Under the ROC curve (AUC). This enables summarising the trade-off between the true and false positive rates of a model by using different probability thresholds. As shown in Fig. \ref{fig::fig4}, our network (DenseLSTM) achieved the highest AUC values of 0.9972 and 0.9760, for the mid-term and final-term examinations, respectively. The LSTM and RNN systems also demonstrated a similar performance for the mid-term examination by achieving AUC values of 0.9870 and 0.9810, respectively. However, the AUC values achieved by both these approaches for the final-term examinations were lower than 0.94. These comparisons demonstrate that our network has outperformed most of the other benchmark networks for classifying normal and abnormal behaviours of students at online examinations with low sensitivity and high specificity.

\begin{table}[!t]
\renewcommand{\arraystretch}{1.2}
\caption{Performance evaluation of the mid-term and final-term online examinations and overall. The highest accuracy figures obtained are highlighted in bold font.}
\label{tab::tab3}
\centering
\begin{tabular}{lccc}
\hline
\hline
\textbf{Networks} & \textbf{Mid-term (\%)} & \textbf{Final-term (\%)} & \textbf{Overall (\%)} \\ \hline
DNN & 82.74 & 52.68 & 67.71 \\
LSTM & 94.49 & 89.29 & 91.89 \\
RNN & 87.20 & 85.02 & 86.11 \\ 
\textbf{DenseLSTM} & \textbf{97.77} & \textbf{92.86} & \textbf{95.32} \\ 
\hline
\hline
\end{tabular}
\end{table}

\subsection{Discussion}
Our experimental analysis and results demonstrate that the proposed approach can successfully address the challenges pertaining to cheating at online examinations, and in preventing such abnormal behaviour of students. The proposed intelligence agent framework utilises the IP detector and the DenseLSTM network, which enable monitoring the students through network protocols and behaviour analysis to identify any potential plagiarism.

As the examinations were evaluated over two terms, our results have shown a consistently high accuracy. The results confirm the performance of the proposed network to be superior in detecting abnormal behaviour at online examinations due to its ability to better maintain ``collective knowledge'' from the extracted features within the subsequent layers, instead of ``forgetting'' the information. This supports our assumption that DenseLSTM performs better than other neural networks.

\section{Conclusion}
Online learning is a new and exciting opportunity for students and education institutions, which is gaining momentum. In today’s environment, e-learning presents unique opportunities, but also unique challenges. The primary area of concern in online assessments is academic dishonesty in the form of cheating, which students attempt to achieve employing numerous avenues. Therefore, it is the responsibility of the education institutions to implement more effective measures to detect academically dishonest behaviour. This paper discusses the concerns around online cheating and offers plausible mechanisms of monitoring and curtailing such incidences using AI technology.

We have demonstrated the effectiveness of the proposed e-cheating intelligent agent, which successfully incorporates IP detector and behaviour detector protocols. The agent has been tested on four deep learning algorithms: the DNN, DenseLSTM, LSTM and RNN, using two exam datasets (mid-term and final-term exams). The highest overall accuracy of 95.32\% was achieved by the DenseLSTM. The average accuracy rate is 90\%, which is sufficient to alert the lecturers to review the exam result of concern. We intend to continue developing web-based e-cheating monitoring systems in the future. Such systems would potentially provide a user-friendly interface for tasks such as uploading exam results, choosing algorithms and other features. The system will be tested among lecturers of various subjects, where they would be able to set specific features that they wish to monitor and for detecting abnormal behaviour of students. The existing agent will be improved based on their feedback.



%





\ifCLASSOPTIONcaptionsoff
  \newpage
\fi



\bibliographystyle{IEEEtran}
\bibliography{IEEEabrv, mybib}

\begin{thebibliography}{10}
\providecommand{\url}[1]{#1}
\csname url@samestyle\endcsname
\providecommand{\newblock}{\relax}
\providecommand{\bibinfo}[2]{#2}
\providecommand{\BIBentrySTDinterwordspacing}{\spaceskip=0pt\relax}
\providecommand{\BIBentryALTinterwordstretchfactor}{4}
\providecommand{\BIBentryALTinterwordspacing}{\spaceskip=\fontdimen2\font plus
\BIBentryALTinterwordstretchfactor\fontdimen3\font minus
  \fontdimen4\font\relax}
\providecommand{\BIBforeignlanguage}[2]{{%
\expandafter\ifx\csname l@#1\endcsname\relax
\typeout{** WARNING: IEEEtran.bst: No hyphenation pattern has been}%
\typeout{** loaded for the language `#1'. Using the pattern for}%
\typeout{** the default language instead.}%
\else
\language=\csname l@#1\endcsname
\fi
#2}}
\providecommand{\BIBdecl}{\relax}
\BIBdecl

\bibitem{Sarrayrih2013}
M.~A. Sarrayrih and M.~Ilyas, ``Challenges of online exam, performances and
  problems for online university exam,'' \emph{International Journal of
  Computer Science}, vol.~10, no.~1, pp. 439--443, 2013.

\bibitem{Jalali2017}
K.~Jalali and F.~Noorbehbahani, ``An automatic method for cheating detection in
  online exams by processing the student's webcam images,'' in \emph{Proc. 3rd
  Conference on Electrical and Computer Engineering Technology (E-Tech 2017)},
  Tehran, Iran, 2017, pp. 1--6.

\bibitem{Raud2019}
Z.~Raud and V.~Vodovozov, ``Advancements and restrictions of e-assessment in
  view of remote learning in engineering,'' in \emph{Proc. 2019 {IEEE} 60th
  International Scientific Conference on Power and Electrical Engineering of
  Riga Technical University (RTUCON)}, Riga, Latvia, 2019, pp. 1--6.

\bibitem{Rowe2004}
N.~C. Rowe, ``Cheating in online student assessment: Beyond plagiarism,''
  \emph{On-Line Journal of Distance Learning Administration}, pp. 1--8, 2004.

\bibitem{Moten2013}
J.~{Moten Jr.}, A.~Fitterer, E.~Brazier, J.~Leonard, and A.~Brown, ``Examining
  online college cyber cheating methods and prevention measures,'' \emph{The
  Electronic Journal of e-Learning}, vol.~11, no.~2, pp. 139--146, 2013.

\bibitem{Chuang2017}
C.~Y. Chuang, S.~D. Craig, and J.~Femiani, ``Detecting probable cheating during
  online assessments based on time delay and head pose,'' \emph{Higher
  Education Research \& Development}, vol.~36, no.~6, pp. 1123--1137, 2017.

\bibitem{ourdb2020}
\BIBentryALTinterwordspacing
7wiseup behaviour dataset. [Online]. Available:
  \url{http://7wiseup.com/research/e-cheating/data/}
\BIBentrySTDinterwordspacing

\bibitem{Roger2006}
C.~F. Roger, ``Faculty perceptions about e-cheating during online testing,''
  \emph{Journal of Computing Sciences in Colleges}, vol.~22, no.~2, pp.
  206--212, 2006.

\bibitem{Cluskey2011}
G.~R. Cluskey, C.~R. Ehlen, and M.~H. Raiborn, ``Thwarting online exam cheating
  without proctor supervision,'' \emph{Journal of Academic and Business
  Ethics}, vol.~4, no.~1, pp. 1--7, 2011.

\bibitem{Bella2015}
G.~Bella, R.~Giustolisi, G.~Lenzini, and P.~Y.~A. Ryan, ``A secure exam
  protocol without trusted parties,'' in \emph{Proc. International Conference
  on ICT Systems Security and Privacy Protection}, Hamburg, Germany, 2015, pp.
  495--509.

\bibitem{Ullah2016}
A.~Ullah, H.~Xiao, and T.~Barker, ``A classification of threats to remote
  online examinations,'' in \emph{Proc. {IEEE} 7th Annual Information
  Technology, Electronics and Mobile Communication Conference (IEMCON)},
  Vancouver, BC, Canada, 2016, pp. 1--7.

\bibitem{Ullah2018}
A.~{Ullah}, H.~{Xiao}, and T.~{Barker}, ``A dynamic profile questions approach
  to mitigate impersonation in online examinations,'' \emph{Journal of Grid
  Computing}, vol.~17, no.~2, p. 209–223, 2018.

\bibitem{Fernandes2018}
G.~{Fernandes Jr.} \emph{et~al.}, ``A comprehensive survey on network anomaly
  detection,'' \emph{Telecommunication Systems}, vol.~70, p. 447–489, 2018.

\bibitem{Mariani2012}
L.~Mariani and D.~Micucci, ``Audentes: Automatic detection of te{N}tative
  plagiarism according to a r{E}ference solution,'' \emph{ACM Transactions on
  Computing Education}, vol.~12, no.~2, p. 1–26, 2012.

\bibitem{Chen2014}
K.~Chen, P.~Liu, and Y.~Zhang, ``Achieving accuracy and scalability
  simultaneously in detecting application clones on android markets,'' in
  \emph{Proc. 36th International Conference on Software Engineering},
  Hyderabad, India, 2014, p. 175–186.

\bibitem{Herrera2019}
G.~{Herrera}, M.~{Nuñez-del-Prado}, J.~G.~L. {Lazo}, and H.~{Alatrista},
  ``Through an agnostic programming languages methodology for plagiarism
  detection in engineering coding courses,'' in \emph{Proc. {IEEE} World
  Conference on Engineering Education (EDUNINE)}, Lima, Peru, 2019, pp. 1--6.

\bibitem{Pawelczak2018}
D.~Pawelczak, ``Benefits and drawbacks of source code plagiarism detection in
  engineering education,'' in \emph{Proc. {IEEE} Global Engineering Education
  Conference (EDUCON)}, Tenerife, Spain, 2018, pp. 1048--1056.

\bibitem{Prathish2016}
S.~{Prathish}, S.~{Athi Narayanan}, and K.~{Bijlani}, ``An intelligent system
  for online exam monitoring,'' in \emph{Proc. International Conference on
  Information Science (ICIS)}, Kochi, India, 2016, pp. 138--143.

\bibitem{Narayanan2014}
A.~{Narayanan}, R.~M. {Kaimal}, and K.~{Bijlani}, ``Yaw estimation using
  cylindrical and ellipsoidal face models,'' \emph{{IEEE} Transactions on
  Intelligent Transportation Systems}, vol.~15, no.~5, pp. 2308--2320, 2014.

\bibitem{Wlodarczyk2016}
M.~{Wlodarczyk}, D.~{Kacperski}, P.~{Krotewicz}, and K.~{Grabowski},
  ``Evaluation of head pose estimation methods for a non-cooperative biometric
  system,'' in \emph{Proc. 23rd International Conference Mixed Design of
  Integrated Circuits and Systems (MIXDES)}, Lodz, Poland, 2016, pp. 394--398.

\bibitem{Hu2018}
S.~{Hu}, X.~{Jia}, and Y.~{Fu}, ``Research on abnormal behavior detection of
  online examination based on image information,'' in \emph{Proc. 10th
  International Conference on Intelligent Human-Machine Systems and Cybernetics
  (IHMSC)}, Hangzhou, China, 2018, pp. 88--91.

\bibitem{Mahadi2018}
N.~A. Mahadi \emph{et~al.}, ``A survey of machine learning techniques for
  behavioral-based biometric user authentication,'' \emph{Recent Advances in
  Cryptography and Network Security}, 2018.

\bibitem{Ghizlane2019}
M.~Ghizlane, F.~H. Reda, and B.~Hicham, ``A smart card digital identity check
  model for university services access,'' in \emph{Proc. the 2nd International
  Conference on Networking, Information Systems \& Security}, New York, NY,
  USA, 2019, pp. 1--4.

\bibitem{Ghizlane2019b}
M.~{Ghizlane}, B.~{Hicham}, and F.~H. {Reda}, ``A new model of automatic and
  continuous online exam monitoring,'' in \emph{Proc. 2019 International
  Conference on Systems of Collaboration Big Data, Internet of Things Security
  (SysCoBIoTS)}, Casablanca, Morocco, 2019, pp. 1--5.

\bibitem{Garg2020}
K.~{Garg}, K.~{Verma}, K.~{Patidar}, N.~{Tejra}, and K.~{Patidar},
  ``Convolutional neural network based virtual exam controller,'' in
  \emph{Proc. 4th International Conference on Intelligent Computing and Control
  Systems (ICICCS)}, Madurai, India, 2020, pp. 895--899.

\bibitem{Fayyoumi2014}
A.~Fayyoumi and A.~Zarrad, ``Novel solution based on face recognition to
  address identity theft and cheating in online examination systems,''
  \emph{Advances in Internet of Things}, vol.~4, pp. 5--12, 2014.

\bibitem{Karim2015}
N.~A. Karim and Z.~Shukur, ``Review of user authentication methods in online
  examination,'' \emph{Asian Journal of Information Technology}, vol.~14,
  no.~5, pp. 166--175, 2015.

\bibitem{Bawarith2017}
R.~Bawarith, A.~Basuhail, A.~Fattouh, and S.~Gamalel-Din, ``E-exam cheating
  detection system,'' \emph{International Journal of Advanced Computer Science
  and Applications}, vol.~8, no.~4, pp. 1--6, 2017.

\bibitem{Hadian2019}
S.~G.~A. {Hadian} and Y.~Bandung, ``A design of continuous user verification
  for online exam proctoring on m-learning,'' in \emph{Proc. 2019 International
  Conference on Electrical Engineering and Informatics (ICEEI)}, Bandung,
  Indonesia, 2019, pp. 284--289.

\bibitem{Mathapati2017}
M.~{Mathapati}, T.~S. {Kumaran}, A.~K. {Kumar}, and S.~V. {Kumar}, ``Secure
  online examination by using graphical own image password schemer,'' in
  \emph{Proc. 2017 {IEEE} International Conference on Smart Technologies and
  Management for Computing, Communication, Controls, Energy and Materials
  (ICSTM)}, Chennai, India, 2017, pp. 160--164.

\bibitem{Ramu2013}
T.~Ramu and T.~Arivoli, ``A framework of secure biometric based online exam
  authentication: An alternative to traditional exam,'' \emph{International
  Journal of Scientific and Engineering Research}, vol.~4, no.~11, pp. 52--60,
  2013.

\bibitem{Mungai2017}
P.~K. {Mungai} and R.~{Huang}, ``Using keystroke dynamics in a multi-level
  architecture to protect online examinations from impersonation,'' in
  \emph{Proc. 2017 {IEEE} 2nd International Conference on Big Data Analysis
  (ICBDA)}, Chennai, India, 2017, pp. 160--164.

\bibitem{Ananya2018}
{Ananya} and S.~{Singh}, ``Keystroke dynamics for continuous authentication,''
  in \emph{Proc. 2018 8th International Conference on Cloud Computing, Data
  Science Engineering (Confluence)}, Noida, India, 2018, pp. 205--208.

\bibitem{Almaiah2020}
M.~A. Almaiah, .~A. Al-Khasawneh, and A.~Althunibat, ``Exploring the critical
  challenges and factors influencing the e-learning system usage during
  covid-19 pandemic,'' \emph{Education and Information Technologies}, pp.
  1--20, 2020.

\bibitem{Rajab2020}
M.~H. Rajab, A.~M. Gazal, and K.~Alkattan, ``Challenges to online medical
  education during the covid-19 pandemic,'' \emph{Cureus}, vol.~12, no.~7, pp.
  1--11, 2020.

\bibitem{Adedoyin2020}
O.~B. Adedoyin and E.~Soykan, ``Covid-19 pandemic and online learning: the
  challenges and opportunities,'' \emph{Interactive Learning Environments}, pp.
  1--13, 2020.

\bibitem{Shukor2015}
N.~A. Shukor, Z.~Tasir, and H.~{Van der Meijden}, ``An examination of online
  learning effectiveness using data mining,'' \emph{Procedia - Social and
  Behavioral Sciences}, vol. 172, pp. 555--562, 2020.

\bibitem{Yan2019}
N.~Yan and O.~T.-S. Au, ``Online learning behavior analysis based on machine
  learning,'' \emph{Asian Association of Open Universities Journal}, vol.~14,
  no.~2, pp. 97--106, 2019.

\bibitem{Gonzalez2020}
C.~S. {González-González}, A.~Infante-Moro, and J.~C. Infante-Moro,
  ``Implementation of e-proctoring in online teaching: A study about
  motivational factors,'' \emph{Sustainability}, vol.~12, p. 3488, 2020.

\bibitem{Yao2016}
Y.~{Yao}, L.~{Zhang}, J.~{Yi}, Y.~{Peng}, W.~{Hu}, and L.~{Shi}, ``A framework
  for big data security analysis and the semantic technology,'' in \emph{Proc.
  2016 6th International Conference on IT Convergence and Security (ICITCS)},
  Prague, Czech Republic, 2016, pp. 1--4.

\bibitem{Hochreiter1997}
S.~Hochreiter and J.~Schmidhuber, ``Long short-term memory,'' \emph{Neural
  Computation}, vol.~9, no.~8, p. 1735–1780, 1997.

\bibitem{Huang2019}
G.~{Huang}, Z.~{Liu}, G.~{Pleiss}, L.~{Van Der Maaten}, and K.~{Weinberger},
  ``Convolutional networks with dense connectivity,'' \emph{{IEEE} Transactions
  on Pattern Analysis and Machine Intelligence}, p. 1–1, 2019.

\bibitem{Nair2010}
V.~Nair and G.~E. Hinton, ``Rectified linear units improve restricted boltzmann
  machines,'' in \emph{Proc. 27th International Conference on Machine Learning
  (ICML)}, Haifa, Israel, 2010, pp. 1--8.

\bibitem{TensorFlow}
\BIBentryALTinterwordspacing
Tensorflow. [Online]. Available: \url{https://tensorflow.org}
\BIBentrySTDinterwordspacing

\bibitem{Kingma2015}
D.~P. Kingma and J.~L. Ba, ``Adam: A method for stochastic optimization,'' in
  \emph{Proc. International Conference on Machine Learning (ICML)}, Lille,
  France, 2015, pp. 1--15.

\bibitem{Kriegeskorte2019}
N.~Kriegeskorte and T.~{Golan}, ``Neural network models and deep learning,''
  \emph{Current Biology}, vol.~29, no.~7, pp. R231--R236, 2019.

\bibitem{Greff2017}
K.~{Greff}, R.~K. {Srivastava}, J.~{Koutník}, B.~R. {Steunebrink}, and
  J.~{Schmidhuber}, ``{LSTM}: A search space odyssey,'' \emph{{IEEE}
  Transactions on Neural Networks and Learning Systems}, vol.~28, no.~10, pp.
  2222--2232, 2019.

\bibitem{Rumelhart1986}
D.~{Rumelhart}, G.~{Hinton}, J.~{Koutník}, and R.~{Williams}, ``Learning
  representations by back-propagating errors,'' \emph{Nature}, vol. 323, p.
  533–536, 1986.

\bibitem{Zerkouk2019}
M.~Zerkouk and B.~Chikhaoui, ``Long short term memory based model for abnormal
  behavior prediction in elderly persons,'' in \emph{Proc. International
  Conference on Smart Homes and Health Telematics (ICOST)}, New York, NY, USA,
  2019, pp. 36--45.

\bibitem{Hao2019}
Z.~{Hao}, M.~{Liu}, Z.~{Wang}, and W.~{Zhan}, ``Human behavior analysis based
  on attention mechanism and {LSTM} neural network,'' in \emph{Proc. 2019
  {IEEE} 9th International Conference on Electronics Information and Emergency
  Communication (ICEIEC)}, Beijing, China, 2019, pp. 346--349.

\end{thebibliography}
\end{document}